# Information model for model driven safety requirements management of complex systems

Romaric Guillerm[1,2], Hamid Demmou[1,2], Nabil Sadou[3]

[1] LAAS-CNRS ,
7 avenue du Colonel Roche, F-31077 Toulouse, France
[2] University of Toulouse; UPS, INSA, INP, ISAE; LAAS,
F-31077 Toulouse, France
{guillerm, demmou}@laas.fr,

[3] SUPELEC / IETR
Avenue de la Boulais, F-35511 Cesson-Sevigne
nabil.sadou@supelec.fr

**Abstract.** The aim of this paper is to propose a rigorous and complete design framework for complex system based on system engineering (SE) principles. The SE standard EIA-632 is used to guide the approach. Within this framework, two aspects are presented. The first one concerns the integration of safety requirements and management in system engineering process. The objective is to help designers and engineers in managing safety of complex systems. The second aspect concerns model driven design through the definition of an information model. This model is based on SysML (System Modeling Language) to address requirements definition and their traceability towards the solution and the Verification and Validation (V&V) elements.

## 1 Introduction

Systems engineering processes are becoming more and more critical and complex. A fundamental characteristic of modern systems is their inherent complexity (Bar-yam, 2005). Complexity implies that different parts of the system are interdependent so that changes in one part may have effects on other parts of the system. These complex systems include the emerging of multiple functions or behaviors, that was not possible before, and they are expected to satisfy additional constraints, especially constraints of reliability, safety and security which are specially addressed in this presentation.

Safety of complex systems relies heavily on the emergent properties that result from the complex interdependencies that exist among the involved components, sub-systems or systems and their environments. It is obvious that the safety properties of complex systems must be addressed in an overall study, with a global framework early in the design phase. Weaknesses of the current safety processes can be resumed in the following points [2]:

- Safety analysis involves some degree of intrinsic uncertainty. So, there is a degree of subjectivity in the identification of safety issues.
- Different groups need to work with different views of the system (e.g. systems engineers' view, safety engineers' view). This is generally a benefit but it can be a weakness if the views are not consistent.
- Definition of the safety requirements, their formalization, and their traceability can be ambiguous or not fully considered.
- System models are developed in electronic form, but no use is made of this for Safety/ Reliability analysis. Ideally there should be a common repository of all requirements, design and safety information.

Some of these points are due to the absence of a safety global approach. Indeed, safety must be addressed as global property and safety requirements [3] must be formulated not only in the small (sub-system level) but in the large (system level).

One of SE processes is requirements engineering (RE) [4]. RE is generally considered in the literature as the most critical process within the development of complex systems [5], [6]. The safety requirements engineering are of concern. A common classification proposed for requirements in the literature classifies requirements as functional or non-functional [7]. Functional requirements describe the services that the system should provide, including the behavior of the system in particular situations. Non-functional requirements are related to emergent system properties such as safety attributes and response time or costs. Generally these Non-functional properties cannot be attributed to a single system component. Rather, they emerge as a result of integrating system components. Furthermost, non-functional requirements are also considered as quality requirements, and are fundamental to determine the success of a system.

Requirements engineering can be divided into 2 mains groups of activities [8]:

1. Requirements development: this activity includes the processes of elicitation [9], documentation, analysis and validation of requirements.
2. Requirements management: this activity includes processes of maintainability management, changes management and requirements traceability [3], [10].

In addition to other processes of system engineering which must be concerned by the safety evaluation, requirements engineering is not an exception. Inadequate or misunderstood requirements have been recognized as the major cause of safety-related catastrophes.

The work presented in this paper is divided into two parts. The first part concerns the integration of safety management in system engineering process. The objective is to help engineers in safety management of complex system by proposing a new approach based on system engineering best practices which can be shared between safety and design engineers. The proposed approach is based on system engineering standard *EIA-632*.

The second part presents an information model based on SysML language to address requirements definition and their traceability [3], [10] towards the solution elements and the V&V (Verification and Validation) elements. Safety requirements are integrated on RE activities, including management activities related to maintenance, traceability, and change management.

The paper is structured into five sections. The second section introduces the design framework. The integration approach is presented in the third section. In the fourth section, the information model is proposed for efficient management of safety requirements. The last section gives some conclusions.

## 2  System engineering approach

The development process highlights the necessary activities, their sequencing and the obtained products. Two approaches for system design have been studied and have been defined. The V-cycle approach and its variants and the processes approach. The processes approach is based on the observation that whatever the strategy used to develop a system, these development activities remain the same. The technical processes are based on different activities of system engineering. They are divided into two categories, system definition processes and Verification and Validation (V&V) processes. They are defined by system engineering standards (IEEE 1220, EIA 632, ISO 15288). The processes approach is more flexible than the V-cycle development; it fits better with complex systems. Moreover, the processes vision does not constrain the sequence of development activities in contrast to the development based on a particular development cycle. This difference is another motivation for adopting a processes approach to systems engineering. In this work the process approach is based on EIA-632 SE standard.

### 2.1    System engineering approach

System Engineering is an interdisciplinary approach, which provides concepts to build new applications. It is a collaborative and interdisciplinary process of problems resolution, supporting knowledge, methods and techniques resulting from the sciences and experiment. System engineering is a framework which helps to define the wanted system, which satisfies identified needs and is acceptable for the environment, while seeking to balance the overall economy of the solution on all the aspects of the problem in all the phases of the development and the life of the system. SE concepts are adequate specifically for complex problems; research issues undergone can bring a solution [11]. In System engineering best practice, we have the following chain: Processes → Methods → Tools.

These entities, such as processes, methods and tools, are the conceptual basis of our approach taken from System engineering best practice.  In the first step, the processes can be identified with respect to the accumulated know-how, and can also be taken from standards like the thirteen generic processes proposed in the EIA-632 standard [12] [13]. The second step concerns the methods to be used. The methods can be either developed or existing one but only if it reflects the whole semantics of the process. No taxonomy has yet been developed for corresponding processes and methods. The third step concerns the tools that do not correspond to the processes but the methods; hence in this approach we cannot use a tool to implement a process without first identifying the associated methods.

**2.2   EIA-632 standard**

One famous standard, currently used in the industrial and military fields, is the EIA-632. This standard covers the product life cycle from the needs capture to the transfer to the user. It gives a system engineering methodology trough 13 interacting processes grouped into 5 groups, covering the management issues, the supply/acquisition, design and requirement, realization and verification/validation processes.

# 3 Integration approach

Managing requirements, and specially safety requirements, at the early stages of system development becomes more and more important as system complexity is continuously growing. Safety of complex systems relies heavily on the emergent properties that result from the complex interdependencies that exist among the involved systems or sub systems and their environments. System Engineering (SE) is the ideal framework for the design of complex system. The need for systems engineering arose with the increase in complexity of systems and projects. A system engineering approach to safety starts with the basic assumption that safety proprieties can only be treated adequately in their entirety when taking into account all the involved variables and the relations between the social and the technical aspects [14]. This basis for system engineering has been stated as the principle that a system is more than the sum of its parts. The Safety management must follow all the steps of SE from the requirements definition to the verification and the validation of the system. The starting point of the work presented in this paper is the following note provided in EIA-632 standard:

**Note**: *Standard does not purport to address all safety problems associated with its use or all applicable regulatory requirements. It is the responsibility of the user of this Standard to establish appropriate safety and health practices and to determine the applicability of regulatory limitations before its use* [13].

The next section aims to help designers in addressing safety problems. It describes, briefly, for each process, how the safety will be considered. It illustrates the proposed approach in term of process which must be defined independently to methods and/or tools (other projects are focused on the methods and tools ([15] [16] for example).

**3.1 Integration approach**

The integration of safety must concern all system engineering processes. This paper is focused only on *System Design processes* and *Technical Evaluation processes*. The implementation of the approach consists in identifying and indicating in which way the safety must be considered for each sub-processes of EIA-632. In other words, the sub-processes of EIA-632 standard are translated or refined in terms of safety and included in system design process.

**3.2 System design processes**

The *System Design Processes* are used to convert agreed-upon requirements of the acquirer into a set of realizable products that satisfy acquirer and other stakeholder requirements. The safety requirements must be taken into account in requirements definition process. It allows the formulation, the definition, the formalization and the analysis of these requirements. Then a traceability [5] model must be build to ensure the consideration of the requirements throughout the development cycle of the system.

### 3.2.1 Requirement definition process

The goal of the requirements definition process is to transform the stakeholder requirements into a set of technical requirements. For functional and non-functional requirements, if this distinction is not possible or not relevant at the requirement elicitation process level, the analyzer may do it to categorize requirements.

In the EIA-632 standard, three types of requirements are defined; the *Acquirer Requirements*, the *Other Stakeholder Requirements* and the *System Technical Requirements*.

Concerning *Acquirer Requirement* and *Other Stakeholder Requirements*, the developer shall define a validated set of acquirer requirements for the system, or portion thereof.

Safety requirements, generally, correspond to constraints in the system. It is necessary to identify and collect all constraints imposed by the acquirer to obtain a safe system. A hierarchical organization associates weight to safety requirements, following their criticality. Safety requirements can be derived from certification or quality requirements or can be explicitly expressed by acquirer or other stakeholder.

The developer shall define a validated set of system technical requirements from the validated sets of acquirer requirements and other stakeholder requirements. For safety requirements, the system technical requirements traduce system performances. It consists on defining safety attributes (determine risk tolerability, SIL level, MTBF, MTBR, failure rate for example). Technical requirements can be derived from a preliminary hazard analysis.

Some Standards are available to guide designer to define safety requirements. For example, safety critical systems within the civil aerospace sector are developed subject to the recommendations outlined in ARP4754 [17] and l'ARP-4761 [18]. These standards give guidance on the 'determination' of requirements, including requirements capture, requirements types and derived requirements.

When requirements are defined, it is possible to define some attributes to facilitate their management by, for example, an expression of requirements using SysML. It allows to link requirements to the design solution.

### 3.2.2 Solution Definition Process

The Solution Definition Process is used to generate an acceptable design solution.

For *Logical Solution Representations,* the developer shall define one or more validated sets of logical solution representations that conform with the technical requirements of the system. The recommendation is to use semi formal / formal models for the solution modeling. The use of formal models allows the automation of verification and analysis. In this processes, safety analysis techniques will be used to determine the best logical solution.

The *physical solution representations* are derived from logical solution representation and must respect all requirements, particularly safety requirements. The same safety analysis may be done for the physical solution representation. The same recommendations than for logical solution remain true.

### 3.3 Technical Evaluation Processes

The Technical Evaluation Processes are intended to be invoked by one of the other processes for engineering a system. Four processes are involved: Systems Analysis, Requirements Validation, System Verification and End Products Validation.

#### 3.3.1 System Analysis Process

In system analysis process, the developer shall perform risk analysis to develop risk management strategies, support management of risks and support decision making. The step of risk analysis can generate some safety requirements other than that defined by the acquirer and stakeholder. These new requirements must be taken into account.

#### 3.3.2 Requirements Validation Process

Requirements Validation is critical to successful system product development and implementation. Requirements are validated when it is certain that they describe the input requirements and objectives such that the resulting system products can satisfy them. In this process, a great attention is given to traceability analysis, which allows verifying all the links among Acquirer and Other Stakeholder Requirements, Technical and Derived Technical Requirements, and Logical Solution Representations. Like other requirements, safety requirements must be validated. The validation allows designing safe system.

To facilitate this step, semi-formal solutions, like UML [19] or SysML [20] can be used for good formulation of requirements. Indeed, the diversity of people concerned by the system design project can have limited knowledge concerning the structure of the future system, that's why industry-scale requirement engineering projects are so hard. So the use of UML or SysML with their different diagrams can be helpful.

#### 3.3.3 System Verification Process

The System Verification Process is used to ascertain that the generated system design solution is consistent with its source requirements, in particular, safety requirements. Some traceability models allow defining the procedure of verifying safety requirement. These procedures are planned at the definition of safety requirement. Simulation is a good and current method used to achieve system verification. Other methods like virtual prototyping, model checking and tests can be used.

## 4 Information model

### 4.1 Requirements management

Requirements management is a crucial activity for the success of a project [5]. Indeed, an important number of documents can be produced in the system definition phase. Without requirements management, it seems impossible to ensure the consistency and the quality necessary for success. Statistical studies show that the success or failure of a project depends, on 40%, on the definition and the management of requirements. Requirements management allows to:
- collect requirements and facilitate their expression,
- detect inconsistencies between them,
- validate them,
- manage requirements changes and ensure their traceability.

It must also ensure that each requirement is properly declined, allocated, monitored, satisfied, verifiable, verified and justified. Figure 1 presents an overview of the requirements management of the EIA-632 standard. The proposed information model is inspired from this pattern. We see that *Other stakeholder requirements*, when added to the A*cquirer requirements*, make up a set of stakeholder requirements that are transformed into *system technical requirements*.

The *logical and physical solution representations* are derived from *technical requirements*. *Design solution* and *specified requirements* are defined by completing the *Solution Definition* Process.

### 4.2 Supporting the design

An information model can be used to:
- guide the design,
- manage requirements changes,
- evaluate project progress,
- or simply to help to understand the system development, on the basis of a common and understandable language.

Indeed, modelling is important for the following reasons:
- it is a support for system analysis and design,
- can be used for sharing knowledge,
- it is used to capitalize knowledge.

Modeling allows the transformation of needs into the system definition. In fact, during this transformation, we will gradually go from abstract concepts to a rigorous definition of the system. In modelling, there are 2 separate areas: the problem area and the possible solutions area. At the beginning of the project, the representation of the problem area is more important than the representation of the possible solutions area. During the progress of the design, representation of possible solutions area will be enriched to achieve the strict definition of the system. In parallel the overall representation of the problem area will be enriched to better define the expectations of the system (needs/requirements) until it is stabilized. The transition between the problem domain and the solution domain is a very delicate point of system engineering. It must be expressed by allocating requirements/properties/constraints on possible solutions. These allocations will generate traceability links which are crucial for the system verification and validation steps. We propose an information model that will be compatible with the requirements of the EIA-632 standard, while adding aspects of safety and risk management. We use SysML language to establish this information model thanks to the different available diagrams which make SysML as the language for systems engineering.

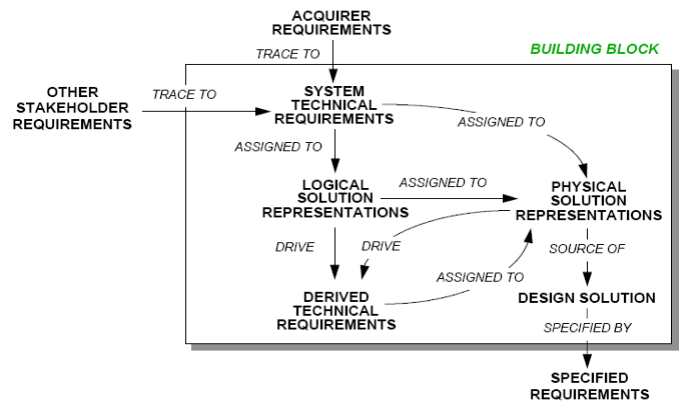

**Fig. 1.** Requirements management of the EIA-632

### 4.3 Requirements modeling and management for safety

*SysML* is a systems modeling language that supports specification, analysis, design, verification and validation of a broad range of complex systems. The language is an evolution of UML 2.0 and is defined for systems that may include hardware, software, information, processes and personnel. It aims to facilitate the communication between heterogeneous teams (mechanical, electrical and software engineers for instance) who work together. The language is effective in specifying requirements, structure, behaviour, allocations of elements to models, and constraints on system properties to support engineering analysis. SysML, through a unique environment integrating requirements, allows modeling and supports different views:
- The requirements: Requirements diagram, Use Case diagram,
- The structure: Block diagram (internal/external),
- The behaviour: Statechart, Activity diagram, Sequence diagram,
- The constraints: Parametric Diagram.

So, SysML seems be an excellent candidate for a common language. It allows sharing specifications of a complex system between different trades. Between design

engineers and safety engineers in our case. Among other benefits of SysML, we can cite:
- Risks identification and creation of a common analytical basis to all participants of the project.
- Facilitates the management of complex projects, the scalability and the maintainability of complex systems.
- Documents and capitalizes the knowledge of all trades in a project.

SysML provides the opportunity to express the requirements using the requirements diagram. It also defines some relationships that link a given requirement to other requirements or elements of the model. It becomes possible to define a hierarchy between requirements, requirement derivation, and requirement satisfaction by a model element, the requirement verification by a test case (*TestCase*) or the requirement refinement. So, this language forms a good basis for our information model. Indeed, in the system definition process, it is necessary to establish a relationship between the identified requirements and the system functions and/or components.

The traceability models linking requirements to the system components allow performing impact analysis of requirements change or modification. Thus, it is possible to assess the consequences of a requirement change on the system safety using the network built between requirements, functions and components.

### 4.4 Proposition

In this part, we propose a system approach to improve requirements management for safe system design. This approach is based on a SysML information model, following the SE process of the EIA-632 standard. This information model is the "system" knowledge basis of the design project, allowing data sharing between all expertise trades (mechanical, hydraulic, thermal, electrical...). Therefore, the model is intended to model the "system" level, showing the interactions with the environment and the connections between the various subsystems.

The information model must be seen as a means of knowledge sharing, including the 3 components: requirements, design solution and V&V. It is considered as the interconnection level between all the different trades.

The safety authorities impose a separation of system design concepts. The requirements, the design solution and the V&V parts must be developed independently. We must be able to distinguish clearly these different concepts.

Based on the previous observation, the proposed approach allows the expression of all the concepts, with a separation between these concepts but with a creation of traceability link between these concepts in order to facilitate understanding and impacts analysis.

With *SysML*, it is easy and possible to mix all the concepts in a single diagram. We propose an extension of *SysML* and information meta-model that allows structuring the elements of the design project with respect of the separation concepts. In other words, our approach allows a rigorous organization of the project design. Indeed, different diagrams manage different concepts.

**4.5 The information model**

The information model (Figure 2) that we propose is adapted to the EIA-632 standard, making a clear distinction between different requirements classes (acquirer, other stakeholders, technical or specified).

To achieve this meta-model in *SysML*, we have extended the language. Firstly, we define new stereotypes to requirements, while adding new attributes to our requirements. Then we define a new link type (*specify*) linking the specified requirements to model elements.

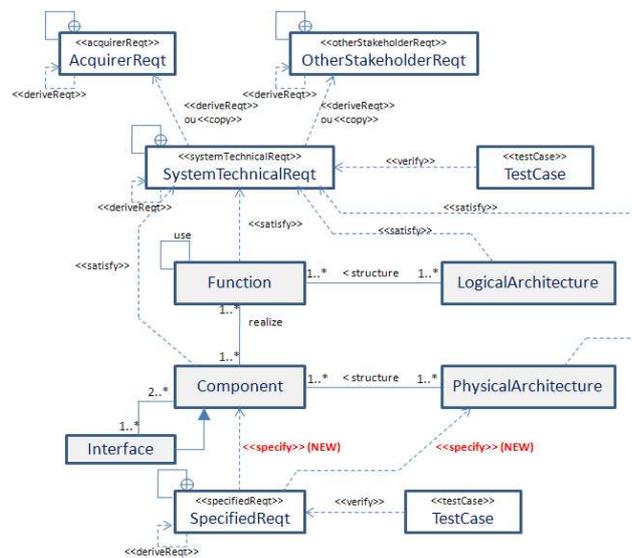

**Fig.2.** Information model

In this model, we have simplified the number of requirements classes. Indeed, we consider that the "*systemTechnicalReqt*" represents the system technical requirements, but also the system technical requirements non-allocated to the logical solution and the derived technical requirements coming from the logical and the physical solutions.

The acquirer and other stakeholders' requirements are represented, knowing that the field 'requirement source' must be consistent with the stereotype and indicates better the concerned stakeholders in the case of "*OtherStakeholderReqt*".

*Note*: We invite readers who are not necessarily friendly with all the above concepts (system technical requirements, logical solution, physical solution …) to refer themselves to the EIA-632 standard [13].

All traceability links requested by the EIA-632 are considered in this model, and the distinction between logical solution (functional part) and physical solution (component part) appears.

In this model, we highlight the definition of interfaces, which are components themselves and which links several components together. The concept of interface is essential for a proper system design. Indeed, it is one source of problems encountered during development.

The last important element that is included in this model, neither a requirement nor a design solution element, is the "*TestCase*". These elements of V&V are included in the model to be directly connected to the requirements they satisfy.

Concerning safety requirements and the consideration of safety in design, which can be derived from risk analysis, a block risk is defined and is linked to safety requirements (see figure 3). In fact, identification of risks is the starting point for many studies about security, but also reliability. Thus, defining a block "risk" in the information model and its link with the safety requirements, allows on one way to improve the system understanding and justifying the requirement, and on the other way to show that all the identified risks are taken into account.

Impact analyses also derive benefit from the presence of risks in the information model, because the risks, which could be challenged by model element (requirement, function, component…) changing, can be viewed directly.

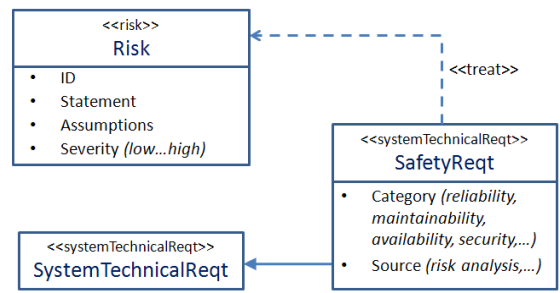

**Fig. 3.** From risk to safety requirements

## 5. Conclusion

Our contribution in this paper can be summarized in the following points: firstly, we illustrate the importance of a global approach for safety evaluation and management. Considering this point we proposed a system engineering approach based on EIA-632 standard. Then we have focused our presentation on the integration of safety management in system engineering processes from requirements definition process to verification and validation process.

Requirements engineering is a crucial activity for the success of a project of complex system design and development. An effective requirements management is necessary. So, in the second part we introduced an information model based on the SysML language. We proposed an extension of this language to define a meta-model by adding new stereotypes for requirements and new attributes to requirements. We also defined a new type of links (*specify*) which link specified requirements to the elements of the model. The proposed model allows the expression of all the handled concepts, and the creation of traceability links between the concepts to facilitate the comprehension and/or the impacts analysis.